\documentclass[preprint,dvipdfmx]{ptephy}

\preprintnumber{KYUSHU-HET-168}

\usepackage{amssymb}
\usepackage{amsthm}
\usepackage{amsmath}
\usepackage{booktabs}
\usepackage{bbm}
\usepackage{mathtools}
\DeclareMathOperator{\tr}{tr}

\DeclareMathOperator{\re}{Re}
\DeclareMathOperator{\im}{Im}

\newcommand{\Slash}[1]{{\ooalign{\hfil/\hfil\crcr$#1$}}}
\numberwithin{equation}{section}

\usepackage{hyperref}
\usepackage{esint}

\begin{document}

\title{A dilaton-pion mass relation}

\author{%
\name{\fname{Aya} \surname{Kasai}}{1},
\name{\fname{Ken-ichi} \surname{Okumura}}{1},
and
\name{\fname{Hiroshi} \surname{Suzuki}}{1,\ast}
}

\address{%
\affil{1}{Department of Physics, Kyushu University, 744 Motooka, Nishi-ku,
Fukuoka, 819-0395, Japan}
\email{hsuzuki@phys.kyushu-u.ac.jp}
}

\begin{abstract}%
Recently, Golterman and Shamir presented an effective field theory which is
supposed to describe the low-energy physics of the pion and the dilaton in an
$SU(N_c)$ gauge theory with $N_f$ Dirac fermions in the fundamental
representation. By employing this formulation with a slight but important
modification, we derive a relation between the dilaton mass squared~$m_\tau^2$,
with and without the fermion mass~$m$, and the pion mass squared~$m_\pi^2$ to
the leading order of the chiral logarithm. This is analogous to a similar
relation obtained by Matsuzaki and~Yamawaki on the basis of a somewhat
different low-energy effective field theory. Our relation reads
$m_\tau^2=m_\tau^2|_{m=0}+KN_f\Hat{f}_\pi^2m_\pi^2/(2\Hat{f}_\tau^2)+%
O(m_\pi^4\ln m_\pi^2)$ with~$K=9$, where $\Hat{f}_\pi$ and~$\Hat{f}_\tau$ are
decay constants of the pion and the dilaton, respectively. This mass relation
differs from the one derived by Matsuzaki and~Yamawaki on the points that
$K=(3-\gamma_m)(1+\gamma_m)$, where $\gamma_m$ is the mass anomalous dimension,
and the leading chiral logarithm correction is~$O(m_\pi^2\ln m_\pi^2)$.
For~$\gamma_m\sim1$, the value of the decay constant~$\Hat{f}_\tau$ estimated
from our mass relation becomes $\sim50\%$ larger than $\Hat{f}_\tau$ estimated
from the relation of Matsuzaki and~Yamawaki.
\end{abstract}
\subjectindex{B01, B31, B36, B44}
\maketitle

\section{Introduction}
\label{sec:1}
The idea that the spontaneous breaking of a (approximate) dilatational or scale
symmetry and the associated (pseudo) Nambu--Goldstone boson (i.e., the dilaton)
play a certain role in elementary particle physics dates back to the
1970's~\cite{Salam:1970qk,Isham:1970gz,Isham:1971dv,Ellis:1970yd,Ellis:1971sa,%
Zumino:1970tu}. For recent investigations, see for example
Refs.~\cite{Bardeen:1985sm,Leung:1985sn,Clark:1986gx,Goldberger:2008zz,%
Fan:2008jk,Vecchi:2010gj,Crewther:2013vea,Crewther:2015dpa} and references
cited therein. This interesting idea has again attracted attention recently as
it might provide a natural understanding on the appearance of a
\emph{flavor-singlet parity-even\/} light meson in the $N_f=8$ $SU(3)$ gauge
theory~\cite{Aoki:2014oha,Appelquist:2016viq}.\footnote{Such a flavor-singlet
parity-even light state has been observed also in an $SU(3)$ gauge theory with
$N_f=2$ Dirac fermions in the symmetric second-rank
representation~\cite{Fodor:2015vwa}. For recent review on lattice study of many
flavor gauge theories, see Refs.~\cite{DeGrand:2015zxa,Giedt:2015alr}.} The
appearance of such a flavor-singlet light scalar meson is extremely interesting
because, combined with the idea of the walking
technicolor~\cite{Holdom:1981rm,Holdom:1984sk,Yamawaki:1985zg,%
Appelquist:1986an,Appelquist:1986tr,Appelquist:1987fc}, the light scalar meson
might be identified with the light Higgs particle. Thus it seems quite
interesting if the lightness of the scalar meson can be understood as a
consequence of the spontaneous symmetry breaking of a flavor-singlet scalar
symmetry: The dilatational symmetry broken by the fermion condensate is a
natural candidate.

It is well-recognized, however, that it is not simple to formulate the
spontaneous breaking of the dilatation symmetry. The Ward--Takahashi relation
associated with the dilatation is almost always intrinsically broken by the
trace or conformal anomaly; this implies that one cannot derive the
corresponding Nambu--Goldstone theorem. The theories in which the dilatation
holds in quantum level, i.e, conformal field theories, do not possess the
dynamical mass scale and thus we do not expect the condensate of an order
parameter. The notion of the spontaneous breaking of the dilatation and the
associated Nambu--Goldstone boson must thus be essentially approximate. If we
know a parameter which controls the validity of an approximate symmetry, such
as the quark mass for the chiral symmetry in QCD, the notion of the spontaneous
symmetry breaking of an approximate symmetry is still quite
useful~\cite{Weinberg:1978kz,Gasser:1983yg,Gasser:1984gg}. For the dilatation,
however, it is not clear at all whether such a useful parameter which controls
the magnitude of the trace anomaly exists or not.

Recently, Golterman and Shamir made an interesting proposal on this
issue~\cite{Golterman:2016lsd}. They take an $SU(N_c)$ gauge theory with $N_f$
Dirac fermions in the fundamental representation. If $N_f$ is within the
so-called conformal window, $N_f^*(N_c)<N_f<(11/2)N_c$, the theory can be
conformal; here, $N_f^*(N_c)=34N_c^3/(13N_c^2-3)$ in the two-loop approximation.
In~Ref.~\cite{Golterman:2016lsd}, the authors consider confining theories in
which $N_f$ is outside the conformal window $N_f<N_f^*(N_c)$ but is very close
to the lower boundary of the window, $N_f\simeq N_f^*(N_c)$. If $N_f$ is very
close to~$N_f^*(N_c)$, the $\beta$-function in the low-energy region at which
the chiral symmetry is spontaneously broken might be regarded as very small and
the Ward--Takahashi relation associated with the dilatation could be regarded
approximately restored; this is the basic idea
of~Ref.~\cite{Golterman:2016lsd}. Further, to introduce a parameter which
controls the ``closeness'' to the window boundary, they consider the Veneziano
limit~\cite{Veneziano:1976wm} in which $N_c\to\infty$ while the ratio
$n_f\equiv N_f/N_c$ is kept fixed. Then $n_f$ may be regarded as a continuous
parameter and the difference~$n_f^*-n_f$, where
$n_f^*\equiv\lim_{N_c\to\infty}N_f^*(N_c)/N_c$, would be used to parametrize the
``smallness'' of the dilatational symmetry breaking in quantum theory.

On the basis of the above idea, in~Ref.~\cite{Golterman:2016lsd}, Golterman
and Shamir formulated an effective field theory which describes the low-energy
physics of the pion and the dilaton in an $SU(N_c)$ gauge theory with $N_f$
flavors. It is then interesting to study consequences of the effective theory
and compare them with results of the lattice simulation for example, to examine
whether the picture of the ``spontaneous dilatational symmetry breaking'' is
physically relevant or not. This is the motivation of the present work.

In the present paper, by employing the formulation
of~Ref.~\cite{Golterman:2016lsd} with an important modification elucidated
in~Sect.~\ref{sec:2.2}, we derive a relation between the dilaton mass
squared~$m_\tau^2$, with and without the fermion mass~$m$, and the pion mass
squared~$m_\pi^2$ to the leading order of the chiral logarithm. This relation is
analogous to a similar relation obtained by Matsuzaki and
Yamawaki in~Ref.~\cite{Matsuzaki:2013eva} on the basis of a somewhat different
low-energy effective theory. Our relation reads
\begin{equation}
   m_\tau^2=m_\tau^2|_{m=0}+K\frac{N_f\Hat{f}_\pi^2}{2\Hat{f}_\tau^2}m_\pi^2
   +O(m_\pi^4\ln m_\pi^2),
\label{eq:(1.1)}
\end{equation}
with $K=9$, where $\Hat{f}_\pi$ and~$\Hat{f}_\tau$ are
decay constants of the pion and the dilaton, respectively. Our mass relation
differs from the one derived by Matsuzaki and~Yamawaki on the points that
$K=(3-\gamma_m)(1+\gamma_m)$, where $\gamma_m$ is the mass anomalous dimension,
and the leading chiral logarithm correction is~$O(m_\pi^2\ln m_\pi^2)$. The
relation in~Ref.~\cite{Matsuzaki:2013eva} has already been utilized to estimate
the dilaton decay constant~$\Hat{f}_\tau$ from lattice data~\cite{Aoki:2014oha}.
For~$\gamma_m\sim1$, the value of the decay constant~$\Hat{f}_\tau$ estimated
from our mass relation becomes $\sim50\%$ larger than $\Hat{f}_\tau$ estimated
from the relation of Matsuzaki and~Yamawaki. We hope that our mass relation
will be examined by lattice simulations in the future in the parameter region
in which the finite volume effect is well under control.

We basically follow the notation of~Ref.~\cite{Golterman:2016lsd}.

\section{Derivation of the mass relation}
\label{sec:2}
\subsection{Microscopic theory}
\label{sec:2.1}
As Ref.~\cite{Golterman:2016lsd}, our microscopic theory is an $SU(N_c)$ gauge
theory with $N_f$ Dirac fermions in the fundamental representation. We assume
the dimensional regularization with the spacetime dimension~$D=4-2\epsilon$,
which will be especially useful in what follows. Thus we set
$S=\int d^Dx\,\mathcal{L}(x)$, where the Lagrangian density is
\begin{equation}
   \mathcal{L}(x)\equiv\frac{1}{4g_0^2}F_{\mu\nu}^a(x)F_{\mu\nu}^a(x)
   +\Bar{\psi}(x)(\Slash{D}+m_0)\psi(x),
\label{eq:(2.1)}
\end{equation}
where $\Slash{D}=\gamma_\mu D_\mu$ and~$D_\mu=\partial_\mu+A_\mu$.

To constrain the possible form of the low-energy effective theory, following
Ref.~\cite{Golterman:2016lsd} (see also Ref.~\cite{Chacko:2012sy}), we
introduce spurious fields, $\chi(x)$ which is an $N_f\times N_f$ matrix field
and $\sigma(x)\in\mathbbm{R}$, corresponding to the chiral symmetry and the
dilatational symmetry, respectively. The action is thus modified as
\begin{align}
   S&=\int d^Dx\,e^{(D-4)\sigma(x)}
   \biggl\{
   \frac{1}{4g_0^2}F_{\mu\nu}^a(x)F_{\mu\nu}^a(x)
\notag\\
   &\qquad\qquad\qquad\qquad\qquad{}
   +\Bar{\psi}(x)\Slash{D}\psi(x)
   +\Bar{\psi}(x)\left[\chi(x)P_R+\chi(x)^\dagger P_L\right]
   \psi(x)\biggr\},
\label{eq:(2.2)}
\end{align}
so that it is invariant under the $SU(N_f)_L\times SU(N_f)_R$ chiral
transformation and the dilatation. The former is given by
($g_L$, $g_R\in SU(N_f)$)
\begin{align}
   \psi(x)&\to(g_RP_R+g_LP_L)\psi(x),&
   \Bar{\psi}(x)&\to\Bar{\psi}(x)(P_Lg_R^\dagger+P_Rg_L^\dagger),
\notag\\
   \chi(x)&\to g_L\chi(x)g_R^\dagger,&&
\label{eq:(2.3)}
\end{align}
and other fields are kept intact. The latter is realized by
($\lambda\in\mathbbm{R}_+$)
\begin{align}
   A_\mu(x)&\to\lambda A_\mu(\lambda x),
\notag\\
   \psi(x)&\to \lambda^{3/2}\psi(\lambda x),&
   \Bar{\psi}(x)&\to\lambda^{3/2}\Bar{\psi}(\lambda x),
\notag\\
   \chi(x)&\to\lambda \chi(\lambda x),&&
\notag\\
   \sigma(x)&\to\sigma(\lambda x)+\ln\lambda.&&
\label{eq:(2.4)}
\end{align}
These symmetries are of course spurious and, going back to the original
theory~\eqref{eq:(2.1)} by setting,
\begin{equation}
   \sigma(x)=0,\qquad\chi(x)=m_0\mathbbm{1},
\label{eq:(2.5)}
\end{equation}
the symmetries are explicitly broken. Still, these spurious symmetries are
quite useful to determine the possible form of the corresponding effective
theory.

Now, one of the crucial relations for our argument is
\begin{equation}
   \left.\delta_{\chi(x)}S\,\right|_{\sigma(x)=0,\chi(x)=m_0\mathbbm{1}}
   =m_0\Bar{\psi}(x)\psi(x),
\label{eq:(2.6)}
\end{equation}
where we have introduced the notation
\begin{equation}
   \delta_{\chi(x)}
   \equiv
   \re\chi_{ij}(x)\frac{\delta}{\delta\re\chi_{ij}(x)}
   +\im\chi_{ij}(x)\frac{\delta}{\delta\im\chi_{ij}(x)}.
\label{eq:(2.7)}
\end{equation}
In terms of the generating functional of connected correlation functions~$W$
corresponding to~$S$~\eqref{eq:(2.2)}, Eq.~\eqref{eq:(2.6)} says that
\begin{equation}
   \left\langle m_0\Bar{\psi}(x)\psi(x)\right\rangle
   =\left.\delta_\chi(x)W\,\right|_{\sigma(x)=0,\chi(x)=m_0\mathbbm{1}}.
\label{eq:(2.8)}
\end{equation}
In both sides of this expression, we can assume the presence of source fields
for gauge invariant operators.

Another basic relation, which can be obtained from a result
of~Ref.~\cite{Collins:1976yq}, is
\begin{align}
   \left.\frac{\delta}{\delta\sigma(x)}
   S\,\right|_{\sigma(x)=0,\chi(x)=m_0\mathbbm{1}}
   &=(D-4)\mathcal{L}(x)
\notag\\
   &\stackrel{D\to4}{\to}-\partial_\mu S_\mu(x)-m_0\Bar{\psi}(x)\psi(x),
\label{eq:(2.9)}
\end{align}
which holds in correlation functions containing gauge invariant operators only.
We note that the combination $m_0\Bar{\psi}(x)\psi(x)$ is ultraviolet finite.
In this expression, $S_\mu(x)$ denotes the dilatation current, defined by
\begin{equation}
   S_\mu(x)\equiv x_\nu
   \left[T_{\mu\nu}(x)
   +\frac{D-1}{D}\delta_{\mu\nu}
   \Bar{\psi}\left(\frac{1}{2}\overleftrightarrow{\Slash{D}}+m_0\right)
   \psi(x)\right]
\label{eq:(2.10)}
\end{equation}
from the energy--momentum tensor, whose definition
in~Ref.~\cite{Collins:1976yq} is
\begin{align}
   T_{\mu\nu}(x)
   &\equiv\frac{1}{g_0^2}\left[
   F_{\mu\rho}^a(x)F_{\nu\rho}^a(x)
   -\frac{1}{4}\delta_{\mu\nu}F_{\rho\sigma}^a(x)F_{\rho\sigma}^a(x)
   \right]
\notag\\
   &\qquad{}
   +\frac{1}{4}
   \Bar{\psi}(x)\left(\gamma_\mu\overleftrightarrow{D}_\nu
   +\gamma_\nu\overleftrightarrow{D}_\mu\right)\psi(x)
   -\delta_{\mu\nu}\Bar{\psi}(x)
   \left(\frac{1}{2}\overleftrightarrow{\Slash{D}}
   +m_0\right)\psi(x),
\label{eq:(2.11)}
\end{align}
where
\begin{equation}
   \overleftrightarrow{D}_\mu\equiv D_\mu-\overleftarrow{D}_\mu,\qquad
   \overleftarrow{D}_\mu\equiv\overleftarrow{\partial}_\mu-A_\mu.
\label{eq:(2.12)}
\end{equation}
The last term in~Eq.~\eqref{eq:(2.10)}, which is proportional to the equation
of motion of the fermion fields, is added so that $\partial_\mu S_\mu(x)$
generates the dilatation transformation on the fermion fields with the scaling
dimension~$(D-1)/2$ through the Ward--Takahashi (WT) relation.

From Eqs.~\eqref{eq:(2.6)} and~\eqref{eq:(2.9)}, assuming the limit~$D\to4$,
we infer that
\begin{equation}
   \partial_\mu S_\mu(x)
   =-\left.\left[
   \frac{\delta}{\delta\sigma(x)}
   +\delta_{\chi(x)}\right]
   S\,\right|_{\sigma(x)=0,\chi(x)=m_0\mathbbm{1}},
\label{eq:(2.13)}
\end{equation}
or, in terms of the generating functional of the connected correlation
functions~$W$,
\begin{equation}
   \left\langle\partial_\mu S_\mu(x)\right\rangle
   =-\left.\left[
   \frac{\delta}{\delta\sigma(x)}
   +\delta_{\chi(x)}\right]
   W\,\right|_{\sigma(x)=0,\chi(x)=m_0\mathbbm{1}}.
\label{eq:(2.14)}
\end{equation}
This is a fundamental relation for our argument.

\subsection{Low-energy effective theory}
\label{sec:2.2}
Next, we consider a low-energy effective field theory along the line of
reasoning in~Ref.~\cite{Golterman:2016lsd}. We assume that the low-energy
degrees of freedom are represented by the pion field~$\Sigma(x)\in SU(N_f)$ and
by the dilaton field~$\tau(x)\in\mathbbm{R}$. Under the
$SU(N_f)_L\times SU(N_f)_R$ chiral transformation, these field are transformed
as 
\begin{equation}
   \Sigma(x)\to g_L\Sigma(x)g_R^\dagger,\qquad
   \tau(x)\to\tau(x),
\label{eq:(2.15)}
\end{equation}
and, under the dilatation,
\begin{equation}
   \Sigma(x)\to\Sigma(\lambda x),\qquad
   \tau(x)\to\tau(\lambda x)+\ln\lambda.
\label{eq:(2.16)}
\end{equation}
Thus, remembering the transformation laws of the spurion fields
in~Eqs.~\eqref{eq:(2.3)} and~\eqref{eq:(2.4)}, the most general form of an
invariant action to the order~$p^2\sim m$ is given
by~\cite{Golterman:2016lsd}\footnote{If we require the dilatation invariance
in $D$~dimensions, the Lagrangian must be multiplied by the
factor~$e^{-2\epsilon\tau(x)}$, where $D=4-2\epsilon$. This ``evanescent
factor'' contributes, through ultraviolet divergences, from the one-loop order;
its effect on the mass relation is however $O(m_\pi^4)$ and is higher order in
our present treatment. It appears interesting to study a possible effect of
this evanescent factor.}
\begin{align}
   \Tilde{S}&=\int d^Dx\,\biggl\{
   \frac{f_\pi^2}{4}V_\pi(\tau(x)-\sigma(x))e^{2\tau(x)}
   \tr\left[\partial_\mu\Sigma(x)^\dagger\partial_\mu\Sigma(x)\right]
\notag\\
   &\qquad\qquad\qquad{}
   +\frac{f_\tau^2}{2}V_\tau(\tau(x)-\sigma(x))e^{2\tau(x)}
   \partial_\mu\tau(x)\partial_\mu\tau(x)
\notag\\
   &\qquad\qquad\qquad\qquad{}
   -\frac{f_\pi^2B_\pi}{2}V_M(\tau(x)-\sigma(x))e^{y\tau(x)}
   \tr\left[Z_m\chi(x)^\dagger\Sigma(x)+\Sigma(x)^\dagger Z_m\chi(x)\right]
\notag\\
   &\qquad\qquad\qquad\qquad\qquad{}
   +f_\tau^2B_\tau V_d(\tau(x)-\sigma(x))e^{4\tau(x)}
   \biggr\}.
\label{eq:(2.17)}
\end{align}
In this expression, the functions $V_I(\tau)$ ($I=\pi$, $\tau$, $M$, and~$d$)
cannot be determined from the invariance of the action
alone~\cite{Golterman:2016lsd}. Here, we have assumed that the action is
polynomial in the spurion field~$\chi(x)$. Otherwise, the term such
as~$V_X(\tau(x)-\sigma(x))e^{3\tau(x)}\tr[\chi(x)^\dagger\chi(x)]^{1/2}$ must
be taken into account.

In Eq.~\eqref{eq:(2.17)}, we have multiplied the bare spurious field~$\chi(x)$
by the mass renormalization factor~$Z_m$, defined by (we set $D=4-2\epsilon$)
\begin{equation}
   m\equiv Z_m(g)m_0,\qquad g_0^2\equiv\mu^{2\epsilon}g^2Z(g),
\label{eq:(2.18)}
\end{equation}
where $m$ and~$g$ are the renormalized mass and gauge coupling, respectively
(for definiteness, we have assumed the minimal subtraction (MS) scheme with the
renormalization scale~$\mu$), so that
$Z_m\chi(x)=Z_mm_0\mathbbm{1}=m\mathbbm{1}$ becomes a ultraviolet finite
quantity. Note that in our renormalization scheme~\eqref{eq:(2.18)}, the
renormalization constant~$Z_m$ is independent of the spurious field~$\sigma(x)$.
Then, for the action~\eqref{eq:(2.17)} to be invariant
under~Eqs.~\eqref{eq:(2.4)} and~\eqref{eq:(2.16)}, the parameter~$y$ in the
third line of~Eq.~\eqref{eq:(2.17)} must be
\begin{equation}
   y=3.
\label{eq:(2.19)}
\end{equation}

Eq.~\eqref{eq:(2.19)} is also required from the equivalence of the effective
theory~\eqref{eq:(2.17)} and the microscopic theory~\eqref{eq:(2.2)}. Consider
the total divergence of the dilatation current $\partial_\mu S_\mu(x)$ in the
effective theory, which must reproduce the relation~\eqref{eq:(2.14)} for the
generating functional~$W$. As computed in~Appendix~D
of~Ref.~\cite{Golterman:2016lsd}, for the action~\eqref{eq:(2.17)}, the Noether
method for the dilatation~\eqref{eq:(2.16)} yields
\begin{align}
   &\left.\partial_\mu S_\mu(x)\right|_{\sigma(x)=0,\chi(x)=m_0\mathbbm{1}}
\notag\\
   &=\frac{f_\pi^2}{4}V_\pi'(\tau(x))e^{2\tau(x)}
   \tr\left[\partial_\mu\Sigma(x)^\dagger\partial_\mu\Sigma(x)\right]
   +\frac{f_\tau^2}{2}V_\tau'(\tau(x))e^{2\tau(x)}
   \partial_\mu\tau(x)\partial_\mu\tau(x)
\notag\\
   &\qquad{}
   -\frac{f_\pi^2B_\pi m}{2}V_M'(\tau(x))e^{y\tau(x)}
   \tr\left[\Sigma(x)+\Sigma(x)^\dagger\right]
   +f_\tau^2B_\tau V_d'(\tau(x))e^{4\tau(x)}
\notag\\
   &\qquad\qquad{}
   +(4-y)\frac{f_\pi^2B_\pi m}{2}V_M(\tau(x))e^{y\tau(x)}
   \tr\left[\Sigma(x)+\Sigma(x)^\dagger\right]
\notag\\
   &=-\left.\left[
   \frac{\delta}{\delta\sigma(x)}
   +(4-y)\delta_{\chi(x)}\right]
   \Tilde{S}\,\right|_{\sigma(x)=0,\chi(x)=m_0\mathbbm{1}},
\label{eq:(2.20)}
\end{align}
where we have used Eq.~\eqref{eq:(2.18)}. This implies
\begin{equation}
   \left\langle
   \left.\partial_\mu S_\mu(x)\right|_{\sigma(x)=0,\chi(x)=m_0\mathbbm{1}}
   \right\rangle
   =-\left.\left[
   \frac{\delta}{\delta\sigma(x)}
   +(4-y)\delta_{\chi(x)}\right]
   W\,\right|_{\sigma(x)=0,\chi(x)=m_0\mathbbm{1}},
\label{eq:(2.21)}
\end{equation}
which coincides with our basic relation~\eqref{eq:(2.14)} if~$y=3$.

Here, we note that in~Ref.~\cite{Golterman:2016lsd}, the parameter~$y$ in the
third line of~Eq.~\eqref{eq:(2.17)} is taken as
\begin{equation}
   y=3-\gamma_m,
\label{eq:(2.22)}
\end{equation}
where $\gamma_m$ is the mass anomalous dimension, defined by
\begin{align}
   \gamma_m(g)
   &\equiv\left.-\mu\frac{\partial}{\partial\mu}\ln m\right|_{g_0,m_0}
\notag\\
   &=\left.-\mu\frac{\partial}{\partial\mu}g\right|_{g_0}
   \frac{d}{dg}\ln Z_m(g).
\label{eq:(2.23)}
\end{align}
The reasoning which leads to~Eq.~\eqref{eq:(2.22)} is elucidated in detail
in a recent reference, Ref.~\cite{Golterman:2016cdd}; to in our language, it
corresponds to a different renormalization scheme which involves the constant
mode of~$\sigma(x)$, $\sigma_0$ through the relations,
\begin{equation}
   \Tilde{m}\equiv Z_m(\Tilde{g})m_0,\qquad g_0^2\equiv
   e^{-2\epsilon\sigma_0}\mu^{2\epsilon}\Tilde{g}^2Z(\Tilde{g}),
\label{eq:(2.24)}
\end{equation}
where the functions $Z_m(g)$ and~$Z(g)$ themselves are identical to those
in~Eq.~\eqref{eq:(2.18)}. Note that the schemes in~Eqs.~\eqref{eq:(2.18)}
and~\eqref{eq:(2.24)} coincide for~$\sigma_0=0$. In this scheme, we have
\begin{align}
   \left.\frac{\partial}{\partial\sigma_0}
   \ln Z_m(\Tilde{g}(e^{\epsilon\sigma_0}\mu^{-\epsilon}g_0))\right|_{g_0,\mu}
   &=\left.-\mu\frac{\partial}{\partial\mu}
   \ln Z_m(\Tilde{g}(e^{\epsilon\sigma_0}\mu^{-\epsilon}g_0))\right|_{\sigma_0,g_0}
\notag\\
   &=\left.-\mu\frac{\partial}{\partial\mu}\Tilde{g}\right|_{\sigma_0,g_0}
   \frac{d}{d\Tilde{g}}\ln Z_m(\Tilde{g})
\notag\\
   &=\gamma_m(\Tilde{g})
\notag\\
   &\stackrel{\epsilon\to0}{\to}\gamma_m(g),
\label{eq:(2.25)}
\end{align}
where we have noted~Eqs.~\eqref{eq:(2.24)} and~\eqref{eq:(2.23)}. This shows
that the mass renormalization factors in the above two schemes are related as
\begin{equation}
   Z_m(\Tilde{g})=e^{\gamma_m(g)\sigma_0}Z_m(g)
\label{eq:(2.26)}
\end{equation}
at~$D=4$. Thus, if we use this scheme, the third line of~Eq.~\eqref{eq:(2.17)}
would become
\begin{align}
   &-\frac{f_\pi^2B_\pi}{2}V_M(\tau(x)-\sigma(x))e^{y\tau(x)}e^{\gamma_m\sigma(x)}
   \tr\left[Z_m(g)\chi(x)^\dagger\Sigma(x)
   +\Sigma(x)^\dagger Z_m(g)\chi(x)\right]
\notag\\
   &=-\frac{f_\pi^2B_\pi}{2}V_M(\tau(x)-\sigma(x))e^{\gamma_m[\tau(x)-\sigma(x)]}
   e^{(y+\gamma_m)\tau(x)}
\notag\\
   &\qquad\qquad\qquad\qquad\qquad\qquad\qquad{}\times
   \tr\left[Z_m(g)\chi(x)^\dagger\Sigma(x)
   +\Sigma(x)^\dagger Z_m(g)\chi(x)\right].
\label{eq:(2.27)}
\end{align}
Here, we have set $\sigma_0\to\sigma(x)$ as it would be justified in the lowest
order of the derivative expansion. Thus, in this scheme, the invariance of the
effective theory under Eqs.~\eqref{eq:(2.4)} and~\eqref{eq:(2.16)}
requires~Eq.~\eqref{eq:(2.22)}.\footnote{It seems difficult, however, to impose
the invariance for~$D\neq4$ in this scheme, because $\gamma_m(\Tilde{g})$
depends on~$\sigma(x)$ for~$D\neq4$.}

The difference between Eqs.~\eqref{eq:(2.19)} and~\eqref{eq:(2.22)} has,
however, no physical relevance at this stage because, as Eq.~\eqref{eq:(2.27)}
shows, the factor~$e^{\gamma_m[\tau(x)-\sigma(x)]}$ that comes from the difference
can be absorbed into the definition of the yet undermined
function~$V_M(\tau(x)-\sigma(x))$. Only when we make a certain choice on~$V_M$,
the difference between Eqs.~\eqref{eq:(2.19)} and~\eqref{eq:(2.22)} does
matter. Here, following the proposal of~Ref.~\cite{Golterman:2016lsd}, we set
\begin{equation}
   V_\pi(\tau)=V_\tau(\tau)=V_M(\tau)=1,\qquad
   V_d(\tau)=c_0+c_1\tau.
\label{eq:(2.28)}
\end{equation}
These might be regarded as the leading order approximation in the Veneziano
limit with the tuning $N_f\to N_f^*(N_c)$, where $N_f^*(N_c)$ is the number of
flavors at the lower boundary of the conformal window~\cite{Golterman:2016lsd}.
Then the crucial question is that which of the representations,
Eq.~\eqref{eq:(2.17)} or Eq.~\eqref{eq:(2.27)}, is more appropriate for the
reasoning which leads to~$V_M=1$. Recalling the basic reasoning
in~Ref.~\cite{Golterman:2016lsd} that the term with the lowest powers
of~$\sigma(x)$ becomes the leading term in the assumed expansion, we think
that the representation~\eqref{eq:(2.17)} is rather consistent with the
choice~$V_M=1$; Eq.~\eqref{eq:(2.27)} has additional dependences on~$\sigma(x)$
even for~$V_M=1$. This completes the explanation on our
choice~Eq.~\eqref{eq:(2.19)} with~$V_M=1$.

Our low-energy effective theory is thus given by (with~Eq.~\eqref{eq:(2.19)})
\begin{align}
   \left.\Tilde{S}\,\right|_{\sigma(x)=0,\chi(x)=m_0\mathbbm{1}}
   &=\int d^Dx\,\biggl\{
   \frac{f_\pi^2}{4}e^{2\tau(x)}
   \tr\left[\partial_\mu\Sigma(x)^\dagger\partial_\mu\Sigma(x)\right]
   +\frac{f_\tau^2}{2}e^{2\tau(x)}
   \partial_\mu\tau(x)\partial_\mu\tau(x)
\notag\\
   &\qquad\qquad\qquad{}
   -\frac{f_\pi^2B_\pi m}{2}e^{y\tau(x)}
   \tr\left[\Sigma(x)+\Sigma(x)^\dagger\right]
\notag\\
   &\qquad\qquad\qquad\qquad{}
   +f_\tau^2B_\tau e^{4\tau(x)}\left[c_0+c_1\tau(x)\right]
   \biggr\}.
\label{eq:(2.29)}
\end{align}
Here and in what follows, the fact that the low-energy constants $f_\pi$,
$f_\tau$, $B_\pi$, and~$B_\tau$ are independent of the fermion mass~$m$ is
crucially important. This follows from the chiral symmetry of the underlying
action~\eqref{eq:(2.17)}. That is, the mass parameter~$m$ can arise only
through the expectation value of the spurion field~$\chi(x)$.

\subsection{Tree level physics}
\label{sec:2.3}
To read off the tree level physics from Eq.~\eqref{eq:(2.29)}, we set
\begin{equation}
   \Sigma(x)=\exp\left[\frac{2\Tilde{\pi}(x)}{f_\pi}\right],\qquad
   \Tilde{\pi}(x)=\Tilde{\pi}^A(x)t^A,\qquad
   \tr(t^At^B)=-\frac{1}{2}\delta^{AB},
\label{eq:(2.30)}
\end{equation}
and expand the action to yield
\begin{align}
   &\left.\Tilde{S}\,\right|_{\sigma(x)=0,\chi(x)=m_0\mathbbm{1}}
\notag\\
   &=\int d^Dx\,\biggl\{
   -e^{2\tau(x)}
   \tr\left[\partial_\mu\Tilde{\pi}(x)\partial_\mu\Tilde{\pi}(x)
   +m_\pi^2(\tau(x))\Tilde{\pi}(x)\Tilde{\pi}(x)\right]
\notag\\
   &\qquad\qquad\qquad{}
   +\frac{f_\tau^2}{2}e^{2\tau(x)}
   \partial_\mu\tau(x)\partial_\mu\tau(x)+V(\tau(x))
\notag\\
   &\qquad\qquad\qquad\qquad{}
   -\frac{2}{3}\frac{1}{f_\pi^2}e^{2\tau(x)}
   \tr\left[
   \Tilde{\pi}(x)^2\partial_\mu\Tilde{\pi}(x)\partial_\mu\Tilde{\pi}(x)
   -\Tilde{\pi}(x)\partial_\mu\Tilde{\pi}(x)\Tilde{\pi}(x)
   \partial_\mu\Tilde{\pi}(x)
   \right]
\notag\\
   &\qquad\qquad\qquad\qquad\qquad{}
   -\frac{1}{3}\frac{m_\pi^2(\tau(x))}{f_\pi^2}e^{2\tau(x)}
   \tr\left[\Tilde{\pi}(x)^4\right]
   +O(\Tilde{\pi}^6)
   \biggr\},
\label{eq:(2.31)}
\end{align}
where
\begin{equation}
   m_\pi^2(\tau)\equiv 2B_\pi me^{(y-2)\tau},
\label{eq:(2.32)}
\end{equation}
and
\begin{equation}
   V(\tau)\equiv
   f_\tau^2B_\tau e^{4\tau}(c_0+c_1\tau)
   -N_ff_\pi^2B_\pi me^{y\tau}.
\label{eq:(2.33)}
\end{equation}

The minimum of the dilaton potential~\eqref{eq:(2.33)} $w$ is given
by~$V'(w)=0$ and
\begin{align}
   w&=v+y\frac{N_ff_\pi^2B_\pi m}{4c_1f_\tau^2B_\tau}e^{(y-4)w}
\notag\\
   &=v+y\frac{N_f\Hat{f}_\pi^2\Hat{B}_\pi m}{4c_1\Hat{f}_\tau^2\Hat{B}_\tau}
   +O(m^2),
\label{eq:(2.34)}
\end{align}
where $v$ is the potential minimum at the chiral limit~$m\to0$,
\begin{equation}
   v=-\frac{1}{4}-\frac{c_0}{c_1},
\label{eq:(2.35)}
\end{equation}
and we have introduced the ``physical'' parameters in the chiral limit,
\begin{equation}
   \Hat{f}_\pi\equiv f_\pi e^v,\qquad\Hat{f}_\tau\equiv f_\tau e^v,\qquad
   \Hat{B}_\pi\equiv B_\pi e^{(y-2)v},\qquad
   \Hat{B}_\tau\equiv B_\tau e^{2v}.
\label{eq:(2.36)}
\end{equation}

Then, from~Eq.~\eqref{eq:(2.31)}, the physical pion mass squared~$m_\pi^2$ is
given by
\begin{equation}
   m_\pi^2=m_\pi^2(w)\equiv 2B_\pi me^{(y-2)w}.
\label{eq:(2.37)}
\end{equation}
Finally, the dilaton mass squared~$m_\tau^2$ is given
by~$V''(w)/(f_\tau^2e^{2w})$ and, by using~$V'(w)=0$, we have
\begin{align}
   m_\tau^2&=4c_1B_\tau e^{2w}
   +y(4-y)\frac{N_ff_\pi^2B_\pi m}{f_\tau^2}e^{(y-2)w}
\notag\\
   &=4c_1B_\tau e^{2w}
   +y(4-y)\frac{N_f\Hat{f}_\pi^2}{2\Hat{f}_\tau^2}m_\pi^2
\notag\\
   &=4c_1\Hat{B}_\tau
   +y(6-y)\frac{N_f\Hat{f}_\pi^2}{2\Hat{f}_\tau^2}m_\pi^2+O(m^2),
\label{eq:(2.38)}
\end{align}
where Eqs.~\eqref{eq:(2.34)} and~\eqref{eq:(2.37)} have been used in the last
equality. Since $4c_1\Hat{B}_\tau$ is independent of the mass parameter~$m$ as
already noted, using Eq.~\eqref{eq:(2.19)}, we obtain the mass
relation~\eqref{eq:(1.1)} with~$K=9$ in the tree level.

It is instructive to see how the derivation of the (tree-level) mass relation
in~Ref.~\cite{Matsuzaki:2013eva} can be understood in the context of the
present low-energy effective theory. The low-energy effective theory
in~Ref.~\cite{Matsuzaki:2013eva} corresponds to~Eq.~\eqref{eq:(2.29)} with the
following particular choice of parameters (in our notation),
\begin{equation}
   c_0=-\frac{1}{16}\frac{m_\phi^2}{B_\tau}
   +y\frac{N_ff_\pi^2B_\pi m}{4f_\tau^2B_\tau},\qquad
   c_1=\frac{1}{4}\frac{m_\phi^2}{B_\tau},
\label{eq:(2.39)}
\end{equation}
where $m_\phi$ is a mass parameter introduced in~Ref.~\cite{Matsuzaki:2013eva}
which is supposed to be independent of the fermion mass~$m$; the parameter~$y$
is given by~Eq.~\eqref{eq:(2.22)}. The second term in~$c_0$
in~Eq.~\eqref{eq:(2.39)}, which depends on the fermion mass~$m$, arises from
the additional term in the action,\footnote{Recall that we eliminated the
possibility of such a term by requiring that the low-energy effective theory is
polynomial in the spurions.}
\begin{equation}
   \int d^Dx\,\frac{1}{4}yf_\pi^2B_\pi e^{4\tau(x)}
   \left\{N_f\tr\left[\chi(x)^\dagger\chi(x)\right]\right\}^{1/2}.
\label{eq:(2.40)}
\end{equation}
In this setup, thus $c_0$ in~Eq.~\eqref{eq:(2.39)}, and consequently $v$
in~Eq.~\eqref{eq:(2.35)} depends on the mass~$m$,
\begin{equation}
   v=-y\frac{N_ff_\pi^2B_\pi m}{f_\tau^2m_\phi^2},
\label{eq:(2.41)}
\end{equation}
and we have to expand also the first term of~Eq.~\eqref{eq:(2.38)} in the
mass~$m$ as
\begin{equation}
   4c_1\Hat{B}_\tau
   =4c_1B_\tau e^{2v}
   =m_\phi^2-y\frac{N_ff_\pi^2}{f_\tau^2}m_\pi^2+O(m^2).
\label{eq:(2.42)}
\end{equation}
Using this in~Eq.~\eqref{eq:(2.38)}, we have
\begin{equation}
   m_\tau^2=m_\phi^2
   +y(4-y)\frac{N_f\Hat{f}_\pi^2}{2\Hat{f}_\tau^2}m_\pi^2+O(m^2).
\label{eq:(2.43)}
\end{equation}
With Eq.~\eqref{eq:(2.22)}, we have Eq.~\eqref{eq:(1.1)}
with~$K=(3-\gamma_m)(1+\gamma_m)$; this reproduces the tree-level mass relation
in~Ref.~\cite{Matsuzaki:2013eva}.

\subsection{One-loop chiral logarithmic corrections}
\label{sec:2.4}
Next, we study the one-loop radiative corrections to the mass
formula~\eqref{eq:(2.38)}. We will consider only the leading-order chiral
log corrections of the form~$m_\pi^2\ln m_\pi^2$ and $m_\pi^4\ln m_\pi^2$ which
would surpass $m_\pi^2$ and~$m_\pi^4$ in the chiral limit~$m\to0$. Since there
is no reason that the dilaton becomes massless as~$m\to0$, in what follows we
will consider only the radiative corrections due to the pion which becomes
massless in the chiral limit.

From Eq.~\eqref{eq:(2.31)}, by the standard method, we have the one-loop
corrections to the effective action as
\begin{align}
   &{\mit\Gamma}^{(1)}
\notag\\
   &=\int d^Dx\,
   \biggl\{\frac{1}{(4\pi)^2}\frac{m_\pi^2}{f_\pi^2}
   \frac{N_f}{3}
   \left[-\frac{1}{\epsilon}+\ln\left(\frac{m_\pi^2}{4\pi}\right)+\gamma-1
   \right]\tr\left[\partial_\mu\Tilde{\pi}(x)\partial_\mu\Tilde{\pi}(x)\right]
\notag\\
   &\qquad\qquad\qquad{}
   +\frac{1}{(4\pi)^2}\frac{m_\pi^2}{f_\pi^2}
   \left(\frac{N_f}{3}-\frac{1}{N_f}\right)
   \left[-\frac{1}{\epsilon}+\ln\left(\frac{m_\pi^2}{4\pi}\right)+\gamma-1
   \right]m_\pi^2\tr\left[\Tilde{\pi}(x)\Tilde{\pi}(x)\right]
\notag\\
   &\qquad\qquad\qquad\qquad{}
   +\frac{1}{(4\pi)^2}m_\pi^2(N_f^2-1)
   \left\{
   \frac{3-y}{2}\left[-\frac{1}{\epsilon}+\ln\left(\frac{m_\pi^2}{4\pi}\right)
   +\gamma\right]
   +\frac{1}{24}(y^2-4y-8)\right\}
\notag\\
   &\qquad\qquad\qquad\qquad\qquad\qquad\qquad\qquad\qquad\qquad\qquad\qquad{}
   \times\partial_\mu\tau(x)\partial_\mu\tau(x)
\notag\\
   &\qquad\qquad\qquad\qquad\qquad{}
   +\frac{1}{(4\pi)^2}\frac{1}{4}(N_f^2-1)
   \left[m_\pi^2(\tau)\right]^2
   \left\{-\frac{1}{\epsilon}+\ln\left[\frac{m_\pi^2(\tau)}{4\pi}\right]
   +\gamma-\frac{3}{2}
   \right\},
\label{eq:(2.44)}
\end{align}
up to terms irrelevant for the corrections to the mass
formula~\eqref{eq:(2.38)} (the function~$m_\pi^2(\tau)$ is defined
in~Eq.~\eqref{eq:(2.37)}). The ultraviolet divergences in this expression are
canceled by appropriate invariant counterterms of~$O(p^4)$ or~$O(m^2)$
(cf.~\cite{Gasser:1983yg}), such as
\begin{align}
   &e^\sigma(x)
   \tr\left[\chi(x)^\dagger\Sigma(x)+\Sigma(x)^\dagger\chi(x)\right]
   \tr\left[\partial_\mu\Sigma(x)^\dagger\partial_\mu\Sigma(x)\right],
\label{eq:(2.45)}
\\
   &e^{2\sigma}(x)
   \left\{\tr\left[\chi(x)^\dagger\Sigma(x)+\Sigma(x)^\dagger\chi(x)\right]
   \right\}^2,
\label{eq:(2.46)}
\\
   &e^\sigma(x)
   \tr\left[\chi(x)^\dagger\Sigma(x)+\Sigma(x)^\dagger\chi(x)\right]
   \partial_\mu\tau(x)\partial_\mu\tau(x),
\label{eq:(2.47)}
\\
   &e^{2\tau(x)}
   \tr\left[\chi(x)^\dagger\chi(x)\right].
\label{eq:(2.48)}
\end{align}
See also the discussion in~Ref.~\cite{Golterman:2016lsd}. The resulting finite
expression then depends on new low-energy constants, the coefficients of those
higher dimensional terms in the action. Still, the coefficients of the
logarithmic factors are invariant under this renormalization. Thus, after the
renormalization, including only the chiral log corrections, we have the
effective action to the one-loop order as ($\Lambda$ is a renormalization
scale)
\begin{align}
   {\mit\Gamma}
   &=\int d^4x\,\Biggl\{
   -e^{2w'}\left[1-\frac{N_f}{3}L(m_\pi^2)\right]
   \tr\left[\partial_\mu\Tilde{\pi}(x)\partial_\mu\Tilde{\pi}(x)\right]
\notag\\
   &\qquad\qquad{}
   -e^{2w'}m_\pi^2(w')
   \left[1-\left(\frac{N_f}{3}-\frac{1}{N_f}\right)L(m_\pi^2)\right]
   \tr\left[\Tilde{\pi}(x)\Tilde{\pi}(x)\right]
\notag\\
   &\qquad\qquad\qquad{}
   +\frac{f_\tau^2}{2}e^{2w'}
   \left[1+(3-y)r\frac{N_f^2-1}{2N_f}L(m_\pi^2)\right]
   \partial_\mu\tau(x)\partial_\mu\tau(x)
\notag\\
   &\qquad\qquad\qquad\qquad{}
   +V(w')
   +\frac{1}{(4\pi)^2}\frac{1}{4}(N_f^2-1)
   m_\pi^4e^{2(y-2)(w'-w)}
   \ln\left(\frac{m_\pi^2}{\Lambda^2}\right)+O(m_\pi^4)\Biggr\},
\label{eq:(2.49)}
\end{align}
where
\begin{equation}
   L(m_\pi^2)\equiv\frac{m_\pi^2}{(4\pi)^2\Hat{f}_\pi^2}
   \ln\left(\frac{m_\pi^2}{\Lambda^2}\right),\qquad
   r\equiv\frac{2N_f\Hat{f}_\pi^2}{\Hat{f}_\tau^2}.
\label{eq:(2.50)}
\end{equation}
$w'$ is the minimum of the dilaton potential which is given by the last line
of~Eq.~\eqref{eq:(2.49)}. For~$m_\pi\to0$, we have
\begin{align}
   w'=w
   -\frac{1}{8c_1\Hat{B}_\tau}(y-2)rm_\pi^2\frac{N_f^2-1}{2N_f}L(m_\pi^2)
   +O(m_\pi^4),
\label{eq:(2.51)}
\end{align}
where $w$ is the minimum of the tree-level potential, Eq.~\eqref{eq:(2.34)}.

Finally, the dilaton mass is given by the second derivative of the potential
with a correction factor arising from the wave function renormalization. Taking
also the correction to the pion mass into account, we find
\begin{align}
   m_\tau^2&=\left.m_\tau^2\right|_{m=0}
   \left[1-(3-y)r\frac{N_f^2-1}{2N_f}L(m_\pi^2)\right]
\notag\\
   &\qquad{}
   +\frac{y(6-y)}{4}rm_\pi^2
   \left\{1-\left[(3-y)r\frac{N_f^2-1}{2N_f}+\frac{1}{N_f}\right]
   L(m_\pi^2)\right\}
\notag\\
   &\qquad\qquad{}
   -(y-2)(5-y)rm_\pi^2\frac{N_f^2-1}{2N_f}L(m_\pi^2)+O(m_\pi^4).
\label{eq:(2.52)}
\end{align}

Now, we notice that the value~\eqref{eq:(2.19)} has a special meaning in view
of~Eq.~\eqref{eq:(2.52)}. When $y=3$, the log correction in the first line
of~Eq.~\eqref{eq:(2.52)} vanishes and the leading log correction becomes
$O(m_\pi^4\ln m_\pi^2)$ as presented in~Eq.~\eqref{eq:(1.1)}. This logarithmic
correction is certainly sub-dominant compared with~$m_\pi^2$ in the sense of
the conventional chiral perturbation theory. On the other hand, if $y\neq3$,
then the leading log correction becomes enhanced to~$O(m_\pi^2\ln m_\pi^2)$ as
the first line of~Eq.~\eqref{eq:(2.52)} which might exceed the tree-level
quantity~$m_\pi^2$ in the second line of~Eq.~\eqref{eq:(2.52)} in the
conventional chiral limit. This inversion of the expansion ordering can happen
because of the presence of the another mass scale, the dilaton
mass~$m_\tau^2|_{m=0}$, which is not small in the conventional chiral expansion.
Although $m_\tau^2|_{m=0}=4c_1\Hat{B_\tau}$ may be regarded as a small quantity
in the new expansion scheme of~Ref.~\cite{Golterman:2016lsd}, this inversion of
the expansion ordering might be troublesome when the relation is applied to fit
to the lattice data for example. The above our observation shows that such
situation does not occur. In this way, we have obtained Eq.~\eqref{eq:(1.1)}
with~$K=9$ including the leading chiral logarithm.

\section{Conclusion}
\label{sec:3}
In the present paper, from the motivation to examine the validity of the
physical picture of the ``spontaneous dilatational symmetry breaking'' in
nearly-conformal $SU(N_c)$ gauge theories with $N_f$ flavors, we derived a
relation among the dilaton, the pion, and the fermion masses in the chiral
limit. We hope that this mass relation will be tested by lattice simulations in
the future. Generalization to theories with fermions in higher dimensional
gauge representations and supersymmetric theories seems interesting.

\section*{Acknowledgements}
We would to thank Shinya Matsuzaki for an explanation on his work and
Shinya Aoki, Yasumichi Aoki, Oliver B\"ar, Maarten Golterman, and Yigal Shamir
for helpful discussions.
This work was supported by JSPS KAKENHI Grant Numbers JP16J02259 (A.~K.)
and~JP16H03982 (H.~S.).

\end{document}